\documentclass[twocolumn,trackchanges]{aastex701}

\begin{document}

\title{Colors of Life in the Clouds:
Biopigments of Atmospheric Microorganisms as a New Signature to Detect Life on Planets like Earth}

\author[orcid=0000-0001-5008-1249,sname='Lígia F. Coelho']{Lígia F. Coelho}
\altaffiliation{Corresponding author}
\altaffiliation{51 Pegasi b Fellow}
\affiliation{Carl Sagan Institute, Cornell University, Ithaca, NY 14850, USA}
\affiliation{Department of Astronomy, Cornell University, Ithaca, NY 14850, USA}
\affiliation{Cornell Center for Astrophysics and Planetary Science, Ithaca, NY 14850 USA}
\email[show]{ lc992@cornell.edu}  

\author[orcid=0000-0002-0436-1802, sname='Lisa Kaltenegger']{Lisa Kaltenegger} 
\affiliation{Carl Sagan Institute, Cornell University, Ithaca, NY 14850, USA}
\affiliation{Department of Astronomy, Cornell University, Ithaca, NY 14850, USA}
\affiliation{Cornell Center for Astrophysics and Planetary Science, Ithaca, NY 14850 USA}
\email{lkaltenegger@astro.cornell.edu}

\author[orcid=0000-0002-5283-4774,sname=William Philpot]{William Philpot}
\affiliation{Carl Sagan Institute, Cornell University, Ithaca, NY 14850, USA}
\affiliation{School of Civil and Environmental Engineering, Cornell University, Ithaca, NY, 14853, USA}
\email{wdp2@cornell.edu}

\author[orcid=0000-0003-4239-6624, sname=Adam J. Ellington]{Adam J. Ellington}
\affiliation{Department of Microbiology and Cell Science, University of Florida, Gainesville, FL 32611, USA}
\email{aellington1994@gmail.com}

\author[orcid=0000-0002-0366-3008,sname=Noelle Bryan]{Noelle Bryan}
\affiliation{Division of Cardiac Surgery, Brigham and Women’s Hospital, Boston, MA, 02115, USA}
\email{ncbryan@mgb.org}

\author[orcid=0000-0003-0239-4640,sname=Stephen Zinder]{Stephen Zinder}
\affiliation{Carl Sagan Institute, Cornell University, Ithaca, NY 14850, USA}
\affiliation{Department of Microbiology, Cornell University, Ithaca, NY 14853, USA}
\email{shz1@cornell.edu}

\author[orcid=0000-0002-9894-7360,sname=Brent C. Christner]{Brent C. Christner}
\affiliation{Department of Microbiology and Cell Science, University of Florida, Gainesville, FL 32611, USA}
\email{xner@ufl.edu}

\begin{abstract}

When Carl Sagan and Ed Salpeter envisioned potential sinkers, floaters, and hunters living in Jupiter's clouds in 1976 \citep{sagan_particles_1976}, the nature of life in Earth's atmosphere remained widely unknown. Decades later, research has revealed a remarkable variety of microorganisms in our atmosphere. However, the spectral features of airborne microbes as biomarkers for detecting atmospheric life remained a mystery. Here, we present the first reflectance spectra of biopigments of atmospheric microorganisms based on laboratory cultivars of seven microbial strains isolated from Earth's atmosphere. We show their distinct UV-resistant biosignatures and their impacts on models of diverse planetary scenarios, using Habitable Worlds Observatory parameters. The reflectance of these biopigments from aerial bacteria creates the means to detect them on other Earth-like planets. It provides a paradigm shift that moves the search for life beyond the surface of a planet to ecosystems in atmospheres and clouds. 

\end{abstract}



\section{Introduction} 

\subsection{Microorganisms living in air}

Viable microorganisms are found in clouds \citep{christner_geographic_2008, santl-temkiv_hailstones_2013} and at altitudes ranging from the clouds to the stratosphere (e.g., at 1.5 km \citealt{santl-temkiv_hailstones_2013}, 38 km  \citealt{bryan_abundance_2019}), with concentrations of about $10^{4}$-$10^{5}$ viable cells per cubed meter. Their abundance and viability will vary with altitude \citep{bryan_abundance_2019, santl-temkiv_microbial_2022, deleon-rodriguez_microbiome_2013}. An estimated 7.6×$10^{23}$–3.5×$10^{24}$ $a^{-1}$ bacterial particles are emitted to our atmosphere annually, mainly from the continents \citep{burrows_bacteria_2009, mayol_long-range_2017}. 

For some of these microorganisms, the atmosphere plays a crucial role in their life cycle, serving as a transport medium to disperse them over large distances. Part of the microorganisms may form an active microbial ecosystem \citep{lappan_atmosphere_2024}. Current technology cannot yet detect microbial life in our atmosphere using remote-sensing equipment \citep{schuerger_science_2018}, mainly because of its low density. 

Until now, microorganisms from Earth’s atmosphere have not been considered for life detection on Earth-like exoplanets. Limited water activity \citep{santl-temkiv_microbial_2022}, energy, and stresses like oxidation seemed to be a general obstacle for detectable aerial biospheres. However, even in extreme environments, bacteria can be abundant, scavenging trace gases like hydrogen for energy and water \citep{ji_atmospheric_2017}. For example, water produced from atmospheric $H_2$ oxidation is sufficient to sustain a bacterium’s water needs, shown experimentally for 15 days \citep{ortiz_multiple_2021}. Organisms from air and rain were shown to oxidize methane \citep{santl-temkiv_viable_2013}, as some organisms isolated from cloud droplets are thought to grow on methanol, formaldehyde, or formic acid \citep{santl-temkiv_hailstones_2013}.  Organisms that can metabolize trace gases (i.e., $H_2$, CO, $CH_4$) are good candidates to survive and even dominate atmospheres \citep{greening_microbial_2022}.

The survival of microorganisms in the atmosphere requires tolerance to UV radiation, particularly above the ozone layer at 35 km \citep{bryan_abundance_2019, santl-temkiv_microbial_2022}, desiccation \citep{bryan_abundance_2019, santl-temkiv_microbial_2022}, oxidation stress \citep{vaitilingom_potential_2013}, and nutrient scarcity when outside of clouds. Organisms isolated from cloud aerosols show biopigments \citep{bryan_abundance_2019, santl-temkiv_microbial_2022}; pigment production has been widely reported in biosurveys of the atmosphere \citep[][and references therein]{santl-temkiv_microbial_2022}. 

Biopigments are synthesized by various organisms, including bacteria, algae, and fungi, and are often crucial for surviving high-incident sunlight exposure, such as that found in Earth’s atmosphere. Many fundamental questions about pigmented organisms and their place in the biosphere remain unanswered, given that colorful biodiversity patterns seem to relate more to the type of environment than to the organisms themselves \citep{coelho_color_2021}. Some pigments contribute remarkable resistance to UV irradiation, e.g., in polar and desert ecosystems \citep{vitek_discovery_2017}, where carotenoids produced by ice and snow biota play crucial roles in both photoprotection and light harvesting \citep{coelho_color_2021}. 

Note that although diverse viable and stress-tolerant microorganisms have been reported in clouds on Earth, whether they are actively metabolizing while residing in the atmosphere remains an open question. But, even if the microorganism is no longer active, their biopigments can provide signatures of life on the surface \citep[e.g.,][]{coelho_color_2021, hegde_surface_2015, deeg_surface_2018, coelho_purple_2024}, and in the atmosphere of exoplanets, especially for water-rich atmospheres. 

Currently, on modern Earth, the residence time for bacterial transport in the stratosphere is estimated to be as long as 2 to 7 years, based on Brewer-Dobson circulation patterns \citep{haenel_reassessment_2015}. The gravitational settling, turbulent mixing below the tropopause, and limited turbulent uplift of surface microorganisms on modern Earth \citep{carotenuto_measurements_2017} prevent microorganisms from reaching high concentrations. Adding, fluctuation of water availability \citep{santl-temkiv_microbial_2022}, UV above the ozone layer, and extreme conditions will also limit their survival. On other planets, however, conditions of increased and more stable levels of bioavailable water in the atmosphere, different vertical mixing, atmospheric density or composition, particle deposition, and slower turnovers could substantially increase their concentration.

Here, we measure biopigments from organisms isolated from cloud aerosols and the atmosphere \citep[e.g.,][]{bryan_abundance_2019}, and show that UV-resistant pigments from microorganisms isolated from Earth's atmosphere can refocus the search for life toward the clouds on Earth-like planets. 

\subsection{Life in the atmospheres of exoplanets}

To consider the range of evolutionary stages on exoplanets \citep[e.g.,][]{omalley-james_vegetation_2018, omalley-james_expanding_2019, kaltenegger_spectral_2007, kaltenegger_high-resolution_2020}, the search for extraterrestrial life focuses on two main approaches: i) searching for atmospheric chemical biosignature pairs \citep{kasting_remote_2014, kaltenegger_how_2017}, such as the combination of oxygen and methane, and ii) searching for surface signatures, such as biopigments of microorganisms or vegetation in the spectrum \citep[e.g.,][]{coelho_color_2021, hegde_surface_2015, deeg_surface_2018, coelho_purple_2024, omalley-james_vegetation_2018, omalley-james_expanding_2019, seager_vegetations_2005}. Machine learning techniques can prioritize promising rocky exoplanet targets using initial identification of water and biological surfaces through color filters \citep[e.g.,][]{pham_color_2021, vannah_informationalentropic_2025}, for upcoming telescopes like the Extremely Large Telescope and concepts like the Habitable Worlds Observatory (HWO).

First detections of atmospheric signatures such as hydrogen,  oxygen, methane, and carbon dioxide for exoplanets made by ground-based \citep[e.g.,][]{jensen_detection_2012, borsa_high-resolution_2021} and space-based \citep[e.g.,][]{bell_methane_2023, jwst_transiting_exoplanet_community_early_release_science_team_identification_2023} telescopes, for some of the more than 6000 detected exoplanets, show diverse atmospheric compositions. Initial observations of rocky planets in the habitable zone have started but are not yet conclusive \citep[e.g.,][]{lim_atmospheric_2023, ducrot_combined_2024}. 

About 60 known rocky planets orbit in the habitable zone of their star \citep{bohl_probing_2025}, with a range of radii and masses. Among those, especially exoplanets
with water-rich atmospheres, are prime targets for the search for life in clouds. On Earth, the troposphere has variable humidity patterns \citep{gettelman_climatology_2006}, making cloud droplets and bioaerosols the medium where atmospheric microorganisms are in higher density \citep{christner_geographic_2008, santl-temkiv_hailstones_2013, temkiv_microbial_2012}, which should apply to other Earth-like planets as well. Aerial life can also interact with the chemistry of the atmosphere. For example, microbial strains isolated from the atmosphere were shown to grow on aldehydes, alcohols, and carboxylic acids at the same rate as Earth’s atmosphere photooxidation \citep{santl-temkiv_hailstones_2013}, having the potential to significantly contribute to the atmospheric changes and chemical dynamics. 

Until now, clouds have mostly been considered obstacles to identifying signs of life because they can obscure both their atmospheric and surface signatures \citep[e.g.,][]{kaltenegger_spectral_2007, kelkar_earth_2025}, thus challenging both current atmosphere retrievals and future imaging. However, the possibility of a pigmented biosphere in the clouds of other Earth-like planets changes that narrative.

\subsection{Signs of UV-resistant life in clouds}

Biopigments produced by microbial life in our atmosphere show what signatures of extraterrestrial life in the air of other planets could look like. To explore this concept, we examined seven microbial strains that were recovered from Earth's stratosphere (see Table~\ref{tab:strain_info}) at altitudes between 21 and 29 km by Bryan et al. \citep{bryan_abundance_2019}, using a high-altitude balloon platform launched from Ft. Sumner, New Mexico. 

All these microbial strains are psychrotolerant (tolerant to cold temperatures) and produce a range of carotenoids. Carotenoids are intense-colored biopigments (400-600 nm) produced by life for protection against radiation, dryness, lack of resources, and temperature \citep{gao_microbial_2011, seel_carotenoids_2020, kimura_thermal-stable_2025}. 

The seven microbial strains are described as follows. Strain L9-4, closely related to \textit{Modestobacter versicolor}, can produce different pigments -- both pink and red carotenoids and black melanin -- for protection against radiation, and expresses pink carotenoids in our measurements. L94A-1, closely related to \textit{Roseomonas vinacea}, shows dark pink pigments. L9-7, closely related to \textit{Micrococcus luteus}, expressed yellow pigments \citep{huang_reclassification_2019}. \textit{Curtobacterium aetherium} L6-1 expresses a yellow/orange pigment and is highly tolerant to desiccation and UV radiation \citep[e.g.,][]{ellington_genetic_2025}. L7-1, closely related to \textit{Massilia niabensis}, is the only specimen of this group that has been initially isolated from aerosols and expresses yellow pigments. L7-5, \textit{Curtobacterium oceanosedimentum}, is extremely resilient to desiccation and UV-C radiation in a vegetative state, even more so than endospores, and shows yellow/orange pigments. L7-7A, closely related to \textit{Noviherbaspirillum soli}, expresses yellow pigments. More information on the microbial strains can be found in Appendix~\ref{sec:appendix_a}.

In this study, we have cultivated and measured the reflection spectra of these seven aerial microorganisms, in dry and wet conditions, producing the first reference of reflectance spectra of the biopigments of aerial microorganisms (see Figure~\ref{fig:general} and Table~\ref{tab:strain_info}, dataset available online on \href{https://doi.org/10.5281/zenodo.17196858}{Zenodo}). 

\section{Methods} 
\subsection{Reflectance measurements}

The strains were isolated by \citet{bryan_abundance_2019} and shipped to Cornell University. At Cornell, we grew and measured the bacteria following \citet{coelho_color_2021} and \citet{hegde_surface_2015}. 
In summary, all strains were cultured in Reasoner’s 2A broth (R2A) media at room temperature. The cultures were grown aerobically up to a stationary phase, which varied from about 24h to 1 week. 

We used an ASD FieldSpec 4 Spectrometer, which covers the wavelength range from 350–2500 nm at intervals of 1 nm, and an ASD integrating sphere to measure the reflectance of the samples. We deposited cultures homogeneously on a 25 mm plain white mixed cellulose ester filter (0.45 mm) using a 10 mL syringe and a filtration system to the point of filter saturation (10 mL ± 1 mL), corresponding to $10^{7}$ and $10^{8}$ cells per filter, as shown by \citet{hegde_surface_2015}. The maximum saturation approach provides an upper limit of signal strength as a baseline spectral reference for planetary modeling while avoiding experimental artifacts of variable cell densities and ensuring reproducibility. 

The same sample was measured at two different times: immediately after being deposited on the filter (wet) and after 1 week of air drying in the dark at room temperature (dry) \citep{coelho_color_2021}. Filters with only culture media were measured as experimental controls, for both wet measurements (fresh medium in the filter) and dry measurements (dry medium in the filter). Reflectance spectra were presented in absolute values and values normalized to the average reflectance between 300 nm and 350 nm.

\subsection{Data generation and simulations}

We used the Exo-Prime II model \citep{madden_high-resolution_2020} to generate high-resolution Earth-like atmospheric spectra with a minimum resolution of $\lambda/\Delta\lambda = 100.000$. Ocean and snow surface albedos were selected from the United States Geological Survey Spectra Library \citep{clark_usgs_2003}, and the single-layer cloud albedo is based on the 20\,mm MODIS model \citep{king_modis_1997, rossow_advances_1999} following previous work \citep{madden_high-resolution_2020}. We resampled the high-resolution spectra of the atmosphere reflectance of an Earth-like atmosphere and the lower-resolution biota (Figure~\ref{fig:general2}), cloud, ocean, and snow albedos using the Python Spectres 2.2.0 package \citep{carnall_spectres_2017}. We resampled the low-resolution biota, cloud, ocean, and snow albedos to R = 140 (comparable to the HWO concept). Our exoplanet models contain reflectance spectra of snowball planets, and ocean worlds with a different surface coverage of cloud biota for atmospheres with 100\% and 50\% clouds.

\section{Results and Discussion} 
\label{sec:Results}
\begingroup
\begin{table*}
    \centering
    \caption{Absorption features (nm) of biopigmetns from wet and dry biota samples. Taxonomy is represented by 16S rRNA gene Sequence Identity.\vspace{-1em}}
    \begin{tabular}{|c|l|c|c|c|}
        \hline
        \textbf{Strain\_ID} & \textbf{Taxonomy} & \textbf{Color} & \textbf{Wavelength wet (nm)} & \textbf{Wavelength dry (nm)} \\ \hline
        L9-4 & \textit{Modestobacter sp.} (98.7\%) & Pink & 420, 470, 490, 520 & 420, 490, 520 \\ \hline
        L9-9A-1 & \textit{Roseomonas sp.} (98.0\%) & Dark pink & 410, 480, 505, 540 & 480, 505, 540 \\ \hline
        L9-7 & \textit{Micrococcus sp.} (98.3\%) & Yellow & 430, 450, 480 & - \\ \hline
        L6-1 & \textit{Curtobacterium aetherium} (100\%) & Yellow/Orange & 490 & 490, 520 \\ \hline
        L7-1 & \textit{Massilia sp.} (99.2\%) & Yellow & 414, 418 & 410 \\ \hline
        L7-5 & \textit{Curtobacterium sp.} (98.3\%) & Yellow & 430, 460, 480 & 430, 460, 480 \\ \hline
        L7-7A & \textit{Noviberspirillum sp.} (99.6\%) & Yellow/Orange & 400, 415, 490 & 415 \\ \hline
    \end{tabular}
    \label{tab:strain_info}
\end{table*}

\begin{figure*}[p]
    \includegraphics[width=\textwidth]{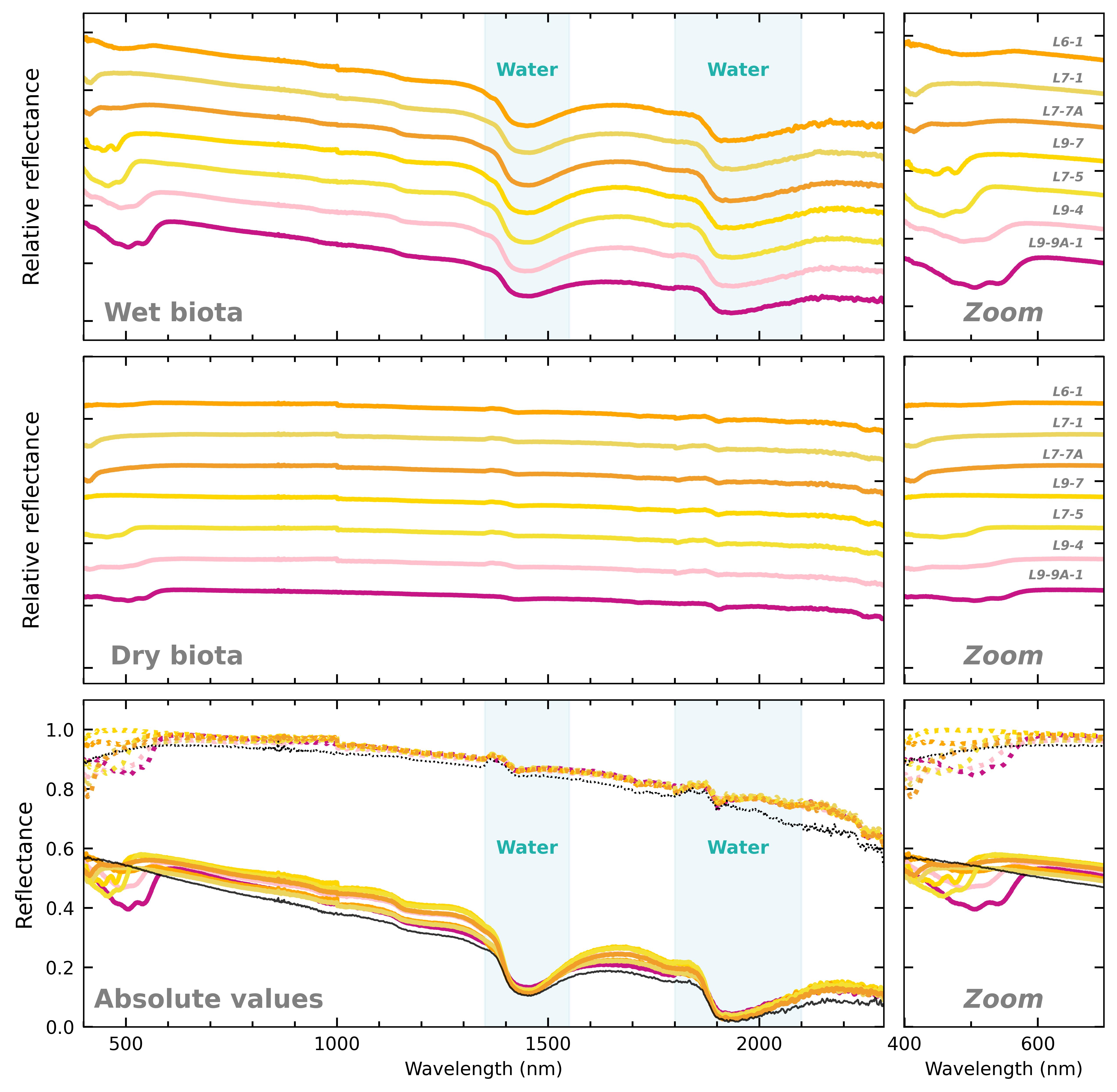}
    \caption{Relative reflectance spectra of seven aerial microorganisms of wet (top panel) and dry (middle panel) measurements, and absolute reflectance (bottom panel - dashed lines for dry biota; solid lines for wet biota) with zoomed-in visible range wavelength panel on the right. Control (black lines in the bottom panel) refers to a filter with culture medium only, where wet control refers to fresh medium and dry control refers to dry medium.  Biopigments show features between 400 and 600 nm (see "Zoom" panels), water absorption features at 1490 nm and 1900 nm are indicated by vertical blue bands. (Data set available on \href{https://doi.org/10.5281/zenodo.17196858}{Zenodo}).}
    \label{fig:general}
\end{figure*}
\endgroup

Looking closely at the individual signatures from the different samples, we see distinctive spectral fingerprints from these microorganisms.
L9-9A-1 reflectance signatures at 410, 480, 505, and 540 nm agree with the absorption pattern of spirilloxanthin \citep{papagiannakis_near-infrared_2003}, a carotenoid known for transferring the photon energy absorbed at 540 nm and dissipating this excess energy as heat \citep{saer_light_2017}. L9-4 and L6-1 spectra show the presence of pink/red/orange carotenoids such as beta-carotene or lycopene \citep[absorption features at 420, 470, 490 nm;][]{sousa_influence_2014}. The yellow pigments from L7-1, L7-7A, L7-5, and L9-7 match with patterns of carotenoids zeaxanthin and lutein described in the literature with absorption features at 410 nm, 440 nm, 430, 450, 460, and 480 nm \citep{nsoukpoe-kossi_absorption_1988}, as well as flavins \citep[absorption features at 450 nm;][]{heelis_photophysical_1982}. Flavins are electron-carrying yellow pigments produced by microorganisms related to several metabolic pathways, usually linked to growth strategies. Note that small wavelength shifts in the literature are likely solvent dependent. We are measuring the whole cell and not the biopigment dissolved in an organic solvent. 

Dry biopigments in Figure~\ref{fig:general} (bottom panel) show an increase in reflectivity \citep{coelho_color_2021}, allowing a better observation of the yellow and fainter pigments from L7-1, L7-7A, and L7-5. Dry measurements are analogs of drier environments than a typical humid ecosystem, such as ice or nonpolar deserts. In these environments, biopigments will reflect strongly (higher intensity), and the increment will be proportional to the water loss \citep{hegde_surface_2015, coelho_purple_2024}. 


The features of the observed carotenoids (absorption features and reflectance peaks – Table~\ref{tab:strain_info}) are more distinct in the wet measurements (Figure~\ref{fig:general}, top panel). This result suggests that atmospheric signatures of carotenoids are easier to distinguish in aerial environments where water is more available because the signature features are more pronounced. However, the overall intensity of the reflectivity of the microorganisms could be higher in drier atmospheres. 

\begin{figure*}[ht!]
\plotone{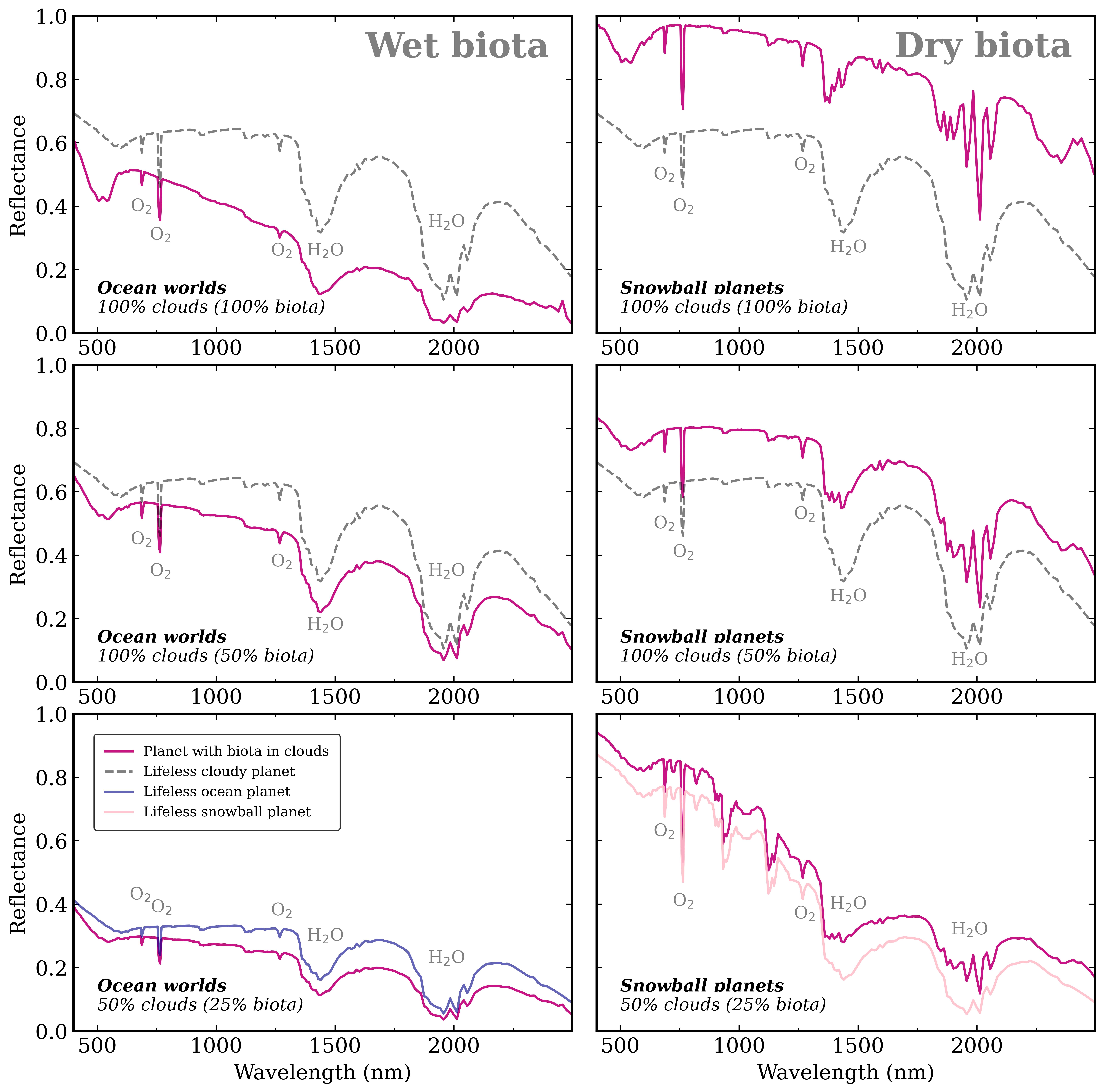}
\caption{Modeled reflectance spectra of different Earth-like planets: ccean worlds (left column), snowball planets (right column), both with clouds covered by strain L9-9A-1 (see Table~\ref{tab:strain_info} and Figure~\ref{fig:general}), resampled for an HWO resolution of R=140. Purple lines represent planets with aerial biota in clouds. Top row: 100\% clouds with 100\% covered by aerial biota. Middle row: 100\% clouds with 50\% covered by aerial biota. Bottom row: 50\% clouds with 50\% covered by aerial biota. Dashed gray lines show ``lifeless cloudy planet'', solid blue shows ``lifeless ocean planet'', and solid pink shows ``lifeless snowball planet'' scenarios. Atmospheric features are labeled O$_2$ at 690, 760, and 1260 nm, and H$_2$O at 1490 and 1900 nm. Individual spectra of albedos are shown in Figure~\ref{fig:appendix}.}
\label{fig:general2}
\end{figure*}

To explore the effect of the biota on the observable reflection spectrum of exoplanets (Figure~\ref{fig:general}), we show the six model scenarios of planets with different surfaces, with resolution resampled to R = 140, comparable to the design concepts of HWO in the early LUVOIR report (Figure~\ref{fig:general2}). We simulate planetary surfaces covered with oceans or snow, a modern Earth-like atmosphere, and varying cloud coverage (100\% and 50\%) with different fractions of clouds covered with biota (100\% and 50\%). Note that especially for fully cloud-covered planets in harsh UV environments like around M stars \citep[e.g.,][]{scalo_m_2007, omalley-james_lessons_2019}, producing pigments for UV protection should be evolutionarily favorable and increase the detectability of life in the atmosphere. While Earth currently is not in such an environment in the solar system, it is unknown whether a cloud-covered planet could show a full or 50\% coverage of biota, which could be detected observationally as shown by the model scenarios (Figure~\ref{fig:general2}). However, while full cloud coverage on an exoplanet generally decreases the possibility of finding surface life through reflection measurements, life in the atmosphere could leave a strong signal, shifting the search for life to the pigments that can be detected in the atmosphere, indicating surface life that is hidden from view.

Adding clouds generally increases the reflectivity and detectability of the exoplanet in the visible compared to an ocean-covered surface because of their higher reflectivity but decreases the overall reflectivity for snowball planets because of the lower reflectivity of clouds compared to snow (see individual albedos in Figure~\ref{fig:appendix}). Once biota is added to the clouds (see Figure~\ref{fig:general2}), it modifies the shape of the reflection spectrum, especially around 500 nm, where the pigment absorption of aerial biota shows strong features for both wet and dry biota (see Figure~\ref{fig:general}). Overall, wet biota reduces the reflectivity of clouds (Figure~\ref{fig:general2} left, ocean worlds), and dry biota increases the reflectivity of clouds (Figure~\ref{fig:general2} right, snowball planets). The interplay between biotic and abiotic components in cloud layers suggests that inhabited clouds would leave a detectable imprint on the planet's overall spectral signature, providing a potential biosignature for future exoplanet observations with missions like HWO. 

While our work represents maximal bioaerosol coverage, future work, including this reference spectra into more complex atmospheric models, could explore minimal cell concentrations for different aerosol distribution assumptions, atmospheric dynamics assumptions, and cloud coverage assumptions. This will be important to assess the sensitivity and limit of detection for different instrument designs.

Aerosolization requires mechanisms such as wind shear, evaporation from liquid surfaces, or wildfire to loft microbes into the atmosphere \citep[e.g.,][]{ellington_dispersal_2024}. On Earth, concentrations have been shown to range from $10^{4}$ to $10^{7}$ cells/m³ in the boundary layer and free troposphere \citep[e.g.,][]{bryan_abundance_2019}, below our experimental saturation levels (see Methods). For exoplanets with denser atmospheres or higher humidity, enhanced vertical mixing could elevate concentrations to detectable thresholds. This implies that detection feasibility increases in dynamic, hydrated environments. 

\section{Conclusion}

Here, we present an additional path for searching for life on Earth-like exoplanets: the search for biopigments as signs of life in clouds. The first reflectance spectra of aerial life demonstrate UV-protective biopigment signatures, offering a critical spectral reference to guide the detection and interpretation of potential biosignatures in the reflected light of Earth-like exoplanets during upcoming missions. This work paves the way for a third paradigm in the search for life on exoplanets, recognizing clouds as surfaces for observable life-supporting ecosystems on Earth-like exoplanets.

\begin{acknowledgments}
LFC is funded by the Heising-Simons Foundation 51 Pegasi b postdoctoral fellowship 2024-5174. LFC, LK, SZ, and WP acknowledge the Carl Sagan Institute at Cornell University. BCC acknowledges funding by NASA’s Exobiology program (80NSSC21K0486) and Interdisciplinary Consortia for Astrobiology Research program (80NSSC23K1477). The authors thank Adam B. Langeveld for his thoughtful comments in the final draft.
\end{acknowledgments}

\begin{contribution}

LFC performed the experiments, measurements, analysis, and modeling, wrote the manuscript, and led the interpretation of the data. LFC and LK conceptualized the research. LK contributed to the modeling, interpretation of the data, and manuscript writing. WP contributed to the measurements, interpretation, and manuscript revision. AJE and NB contributed to the experiments and manuscript revision. SZ and BCC contributed to the experiments, interpretation of data, and manuscript revision.   


\end{contribution}

\clearpage
\appendix
\section{Information on the strains collected in the atmosphere}
\label{sec:appendix_a}

L9-4 is a a closely related strain to \textit{Modestobacter versicolor} (98.7\% 16S rRNA identity), a microorganism that, as the name suggests, produces different pigments, namely both pink/red carotenoids, and melanin (black) for protection against radiation and lack of resources \citep{reddy_modestobacter_2007}. L9-9A-1 is closely related to \textit{Roseomonas vinacea} \citep{zhang_roseomonas_2008} (98.0\% 16S rRNA identity), a dark pink species. L9-7, a strain closely related to \textit{Micrococcus luteus} \citep{huang_reclassification_2019} (98.3\% 16S rRNA identity), which, as the name indicates, is yellow. L6-1 \textit{Curtobacterium aetherium} \citep{mijatovic_scouten_curtobacterium_2025}, and L7-5 \textit{Curtobacterium oceanosedimentum} (98.3\% 16S rRNA identity) are yellow/orange bacteria extremely resilient \citep{reddy_modestobacter_2007} to desiccation and UV-C radiation in the vegetative state, even more so than endospores \citep{bryan_microbial_2017, evseev_curtobacterium_2022}. L7-1 is closely related to \textit{Massilia niabensis} (99.2\% 16S rRNA identity), the only specimens of this group that have been initially isolated from aerosols. Lastly, L7-7A is closely related to the yellow \textit{Noviherbaspirillum soli} \citep{lin_description_2013} (99.6\% 16S rRNA identity). 

\section{Individual albedos used in spectral simulations}
\restartappendixnumbering

Here we show the individual albedos used in Figure~\ref{fig:general2} and described further in the Results and Discussion, section~\ref{sec:Results}.

\begin{figure*}[ht!]
\centering
\includegraphics[width=0.7\textwidth]{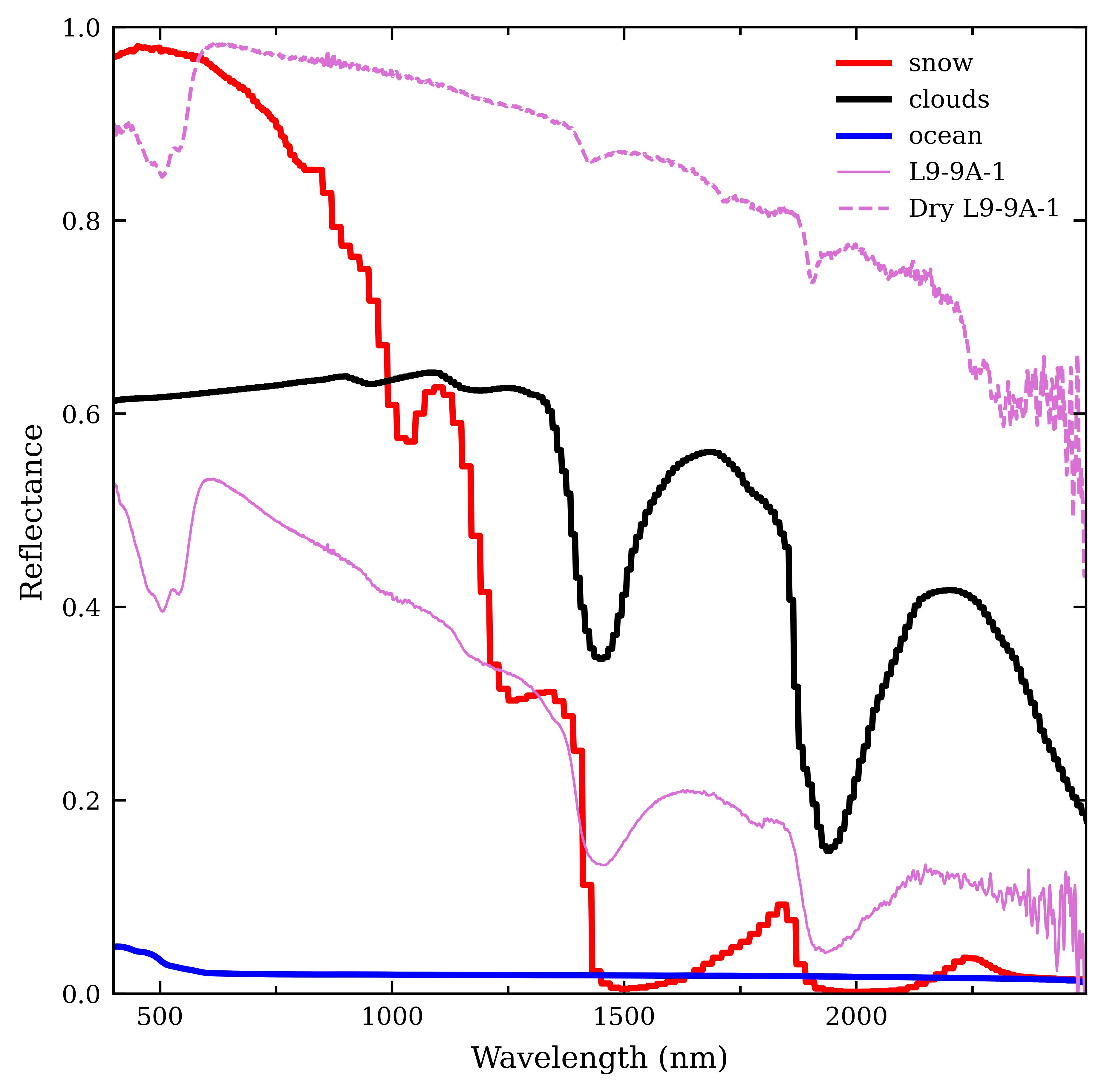}
\caption{Individual spectra of surface albedos used in Figure~\ref{fig:general2}.}
\label{fig:appendix}
\end{figure*}
\clearpage


\bibliography{references2}{}
\bibliographystyle{aasjournalv7}



\end{document}